\documentclass[apj]{emulateapj}
\usepackage{graphicx}
\usepackage{natbib}
\usepackage{times}
\usepackage[colorlinks=true,urlcolor=blue,citecolor=blue,linkcolor=blue]{hyperref}
\newcommand{\ha}{H$\alpha$}

\newcommand{\calong}{\ion{Ca}{2}~H~3968~\AA}
\newcommand{\cah}{\ion{Ca}{2}~H}
\newcommand{\mglong}{\ion{Mg}{2}~2976~\AA} 
\newcommand{\silong}{\ion{Si}{4}~1400~\AA}
\newcommand{\helong}{\ion{He}{2}~304~\AA}
\newcommand{\carbonlong}{\ion{C}{2}~1330~\AA}
\newcommand{\ca}{\ion{Ca}{2}}
\newcommand{\mg}{\ion{Mg}{2}} 
\newcommand{\si}{\ion{Si}{4}}
\newcommand{\he}{\ion{He}{2}}
\newcommand{\carbon}{\ion{C}{2}}

\newcommand{\hinode}{\textit{Hinode}}
\newcommand{\xt}{$xt$}
\newcommand{\kms}{km~s$^{-1}$}
\usepackage{microtype}
\defcitealias{pereira_et_al_2014}{Paper~I}

\begin{document}
\title{On the temporal evolution of spicules observed with IRIS, SDO and {\it Hinode}}
\author{H. Skogsrud\altaffilmark{1}}
\author{L. Rouppe van der Voort\altaffilmark{1}}
\author{B. De Pontieu\altaffilmark{1,2}}
\author{T. M. D. Pereira\altaffilmark{1}}
\affil{\altaffilmark{1}Institute of Theoretical Astrophysics, University of Oslo, P.O. Box 1029 Blindern, N-0315 Oslo, Norway}
\affil{\altaffilmark{2}Lockheed Martin Solar \& Astrophysics Lab, Org.\ A021S, Bldg.\ 252, 3251 Hanover Street Palo Alto, CA~94304 USA}

\begin{abstract}
Spicules are ubiquitous, fast moving jets observed off-limb in chromospheric spectral lines. Combining the recently-launched Interface Region Imaging Spectrograph with the Solar Dynamics Observatory and \hinode, we have a unique opportunity to study spicules simultaneously in multiple passbands and from a seeing free environment. This makes it possible to study their thermal evolution over a large range of temperatures. A recent study showed that spicules appear in several chromospheric and transition region spectral lines, suggesting that spicules continue their evolution in hotter passbands after they fade from \cah. In this follow-up paper we answer some of the questions that were raised in the introductory study. In addition, we study spicules off-limb in \carbonlong\ for the first time. We find that \cah\ spicules are more similar to \mglong\ spicules than initially reported. For a sample of 54 spicules, we find that 44\% of \silong\ spicules are brighter toward the top; 56~\% of the spicules show an increase in \si\ emission when the \cah\ component fades. We find several examples of spicules that fade from passbands other than \cah, and we observe that if a spicule fades from a passband, it also generally fades from the passbands with lower formation temperatures. We discuss what these new, multi-spectral results mean for the classification of type I and II spicules.
\end{abstract}

\section{Introduction}
Spicules have a long and rich scientific history, but their relatively small widths, barely resolved even in modern telescopes, combined with a complex dynamical behavior make spicules a source of lively scientific debate on their origin and role in the solar atmosphere. Traditionally, spicules are observed above the limb in chromospheric spectral lines such as \calong\ and \ha. In the traditional picture, spicules are seen to move outward before falling back toward the limb a few minutes later. Typical lifetimes observed are 2--6~min with typical velocities of 15--40~\kms. \cite{athay_and_holtzer_1982} suggest that the release of gravitational energy from falling spicules could be the source of heating in the atmosphere and \cite{pneuman_and_kopp_1978} indicated that significant spicule material could reach coronal temperatures. \cite{withbroe_1983} studied EUV emission from the downfalling spicule material and indicated that release of gravitational energy is not the primary source of heating in the upper chromosphere. See \cite{beckers_1968,beckers_1972} for reviews of older spicule observations and \cite{sterling_2000} for a review on the theoretical side of early spicule models.

With the emergence of modern space borne telescopes and telescopes utilizing advanced image post-processing, a wealth of new discoveries has been made. \cite{de_pontieu_et_al_2007b} used \cah\ data from the \hinode\ satellite to discover a different class of spicules, most common in quiet Sun regions and coronal hole regions. These so-called type II spicules are predominantly seen to move outward from the limb before fading from the observed passband. Their lifetimes are smaller than what was traditionally observed for spicules, usually less than 2~minutes. The few constraints on the formation of type II spicules, together with their possible vital contribution to the energy budget (due to their sheer number density) have placed type II spicules at the center of many debates, both regarding their very existence and their formation \citep{de_pontieu_et_al_2007b,judge_et_al_2011,zhang_et_al_2012,tiago_et_al_2012,pereira_2013_blur,lipartito_et_al_2014,pereira_et_al_2014}. 

One possible explanation for the fading of \ca\ spicules is heating. \cite{de_pontieu_et_al_2011} show that the disk counterpart of type II spicules can be connected to brightenings in the hotter passbands of the Atmospheric Imaging Assembly \citep[AIA,][]{lemen_et_al_2012} onboard the Solar Dynamics Observatory, which might make the new class of spicules vital in the energy balance of the outer solar atmosphere. \citet[][hereafter \citetalias{pereira_et_al_2014}]{pereira_et_al_2014} use co-observations from three different spacecrafts and find that type II spicules have companion components in the \mglong\ and \silong\ filtergrams of the Interface Region Imaging Spectrograph \citep[IRIS,][]{pontieu_et_al_2014_iris} in addition to \helong\ from AIA. The spicules continue to evolve after fading from the \cah\ passband, strongly suggesting that spicules are the site of heating to at least transition region temperatures. 

Numerical simulations have yet to shed any significant light on this predominantly observational feature, with only one notable exception \citep{martinez-sykora_et_al_2011}. However, numerical experiments will be invaluable in determining in detail the mechanism for formation and heating of spicules. Providing observational constraints will help verification and interpretation of simulation efforts and is one of the goals of this paper. We tackle this issue by looking for answers to the following questions: How common are spicules that disappear from \cah\ and brighten up in \si\ at the same time? Are \mg\ spicules really that different from \ca\ spicules? What fraction of \si\ spicules are brighter near the top? How do the decelerations of type II spicules measured in hotter passbands compare with those reported for type I spicules?

\section{Analysis}
\subsection{The observations}
\label{sec:observations}
\begin{figure*}
  \centering
  \includegraphics[width=0.8\textwidth]{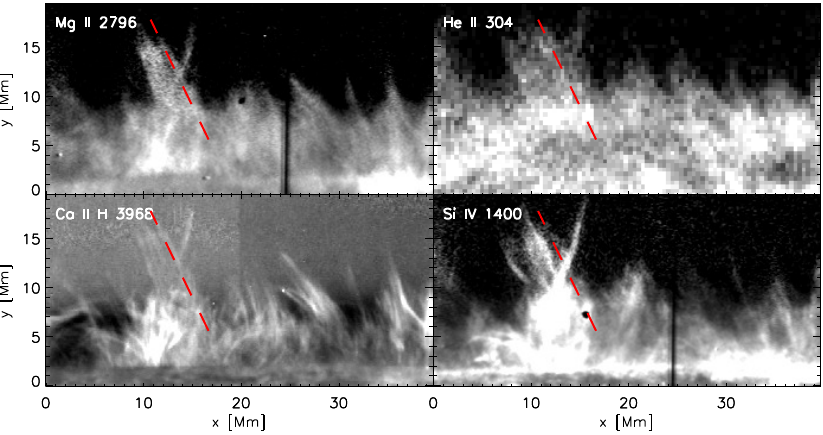}\\
  \includegraphics[width=0.8\textwidth]{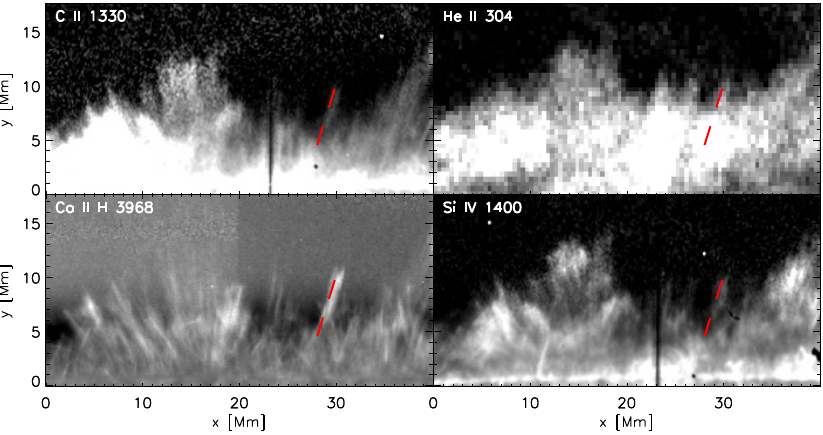}
  \caption{Limb cutouts of the two datasets analyzed. The red dashed line is at the same position in all images of a given dataset. The photospheric limb is nearly parallel to the $x$-axis at $y\sim2$~Mm. The thick vertical black line seen in the IRIS slit-jaw images is the IRIS slit. In the \ca\ images we can see the border between the two CCD sensors as a vertical line at $x\sim20$~Mm because we are pushing the data and looking at slight differences in the dark current correction, which is not perfect. All images except \he\ are radially filtered (see text).}
  \label{fig:fov}
\end{figure*}

In the remainder of this paper the \calong, \mglong, \silong, \carbonlong\ and \helong\ filtergrams are referred to as \ca, \mg, \si, \carbon\ and \he, respectively.

We use coordinated observations from instruments onboard three different space borne telescopes: the Broadband Filter Imager (BFI) on the Solar Optical Telescope (SOT) onboard \hinode\ \citep{sot_overview,sot,hinode_overview}, AIA and IRIS. From \hinode/BFI we analyze filtergrams centered on the \cah\ spectral line. The filter bandwidth is 0.3~nm, the exposure time is 1.8~s and the pixel size is 0$\farcs$11~pix$^{-1}$. From IRIS we analyze the following slit-jaw (SJI) filtergrams: SJI 2796, centered on the line core of \mg, SJI 1330, dominated by the \carbon\ lines, and SJI 1400, dominated by the \si\ lines. The far-UV passbands, \carbon\ and \si, have a bandwidth of 40~\AA, and the near-UV \mg\ has 4~\AA\ band pass. All IRIS slit-jaw images have 8~s exposure time and a pixel size of 0$\farcs$17~pix$^{-1}$. From AIA we use the 304~\AA\ channel which is dominated by \he. The exposure time is 2.9~s and the pixel size is 0$\farcs$6~pix$^{-1}$ and the full width at half maximum (FWHM) of the filter is 12.7~\AA.

Assuming equilibrium conditions we expect the \ca\ and \mg\ passbands to be dominated by radiation from plasma at 10,000 to 15,000~K, \carbon\ from 15,000 to 25,000~K, \si\ from 60,000 to 80,000~K and \he\ from 75,000 to 100,000~K \citep{golding_et_al_2014,carlsson_et_al_2012,pontieu_et_al_2014_iris}.


\begin{deluxetable}{lrrrr}
\tablecaption{IRIS data sets.\label{tab:obs}}
\tablehead{
\colhead{Starting time} & \colhead{($x, y$) Coord.}  & \colhead{$\Delta t$} & \colhead{Duration} & \colhead{Filtergrams} } 
\startdata
2014-02-21T11:24 & $(7\arcsec, -987\arcsec)$ & $19$~s & 90~min & \mg, \si \\
2014-02-19T16:28 & $(6\arcsec, -965\arcsec)$ & $23$~s & 67~min & \carbon, \si
\enddata
\tablecomments{$\Delta t$ is the cadence of the \mg\ or \carbon\ images (for \si\ the cadence was always 19~s). The duration covers the overlapping period between Hinode and IRIS. The first dataset is the same as in \citetalias{pereira_et_al_2014}.}
\end{deluxetable}
\vspace{0.1cm}

We use two different datasets with a different set of IRIS filtergrams available for each, see details in Table~\ref{tab:obs}. The first dataset is the same as in \citetalias{pereira_et_al_2014}. At the time of the observations no coronal hole was present on the South Pole so the region is characterized by quiet Sun conditions. The cadences of BFI/SOT and AIA were 5~s and 12~s, respectively.

We made extensive use of CRISPEX \citep{vissers_et_al_2012} when studying the observations.

\subsection{Aligning the datasets}

The \hinode\ and AIA data are interpolated to the IRIS plate scale. The IRIS data were rotated by 0.6 degrees relative to the center of the slit-jaw images to compensate for the slightly different orientation between IRIS and AIA.

The images observed with the various passbands are different to a varying degree. The alignment of the data from the different spacecrafts poses a formidable challenge because there is no common filtergram or one that is sufficiently similar to be used for unambiguous automatic alignment. Because we are analyzing limb observations, we expect that filtergrams formed at different heights to have a spatial offset in the limb-ward direction. We chose the 1600~\AA\ channel from AIA to serve as a common base for the alignment. The height of formation of the channels is different in magnetic and non-magnetic regions. On-disk in non-magnetic regions, the 1600~\AA, \si\ and \carbon\ channels are dominated by continuum radiation originating from the upper photosphere, and the \ca\ and \mg\ passbands are formed at comparable heights (M. Carlsson, private communication, see also \citealt{martinez-sykora_et_al_2015}). In magnetic regions, however, in particular the \si\ and \carbon\ channels, and to some extent the 1600~\AA\ channel, are dominated by line emission from transition region lines, and formation heights are much less confined. The procedure is then to align IRIS and \hinode\ data to each other indirectly through the 1600~\AA\ channel in non-magnetic regions, because there we expect the channels to be formed at roughly the same height. A precise alignment between the IRIS channels is made possible by the fiducial mark on the slit \citep{pontieu_et_al_2014_iris} and the \he\ passband is aligned to the 1600 channel to within a precision of 1 or 2 AIA pixels by internal alignment with the use of level 1.5 AIA data \citep{lemen_et_al_2012}. 

After various experiments we settled on the following approach: The \si\ to 1600 alignment was done separately for the $x$ and $y$ direction. For the $y$ direction we used cross-correlation on a subsection of the disk field of view (FOV), approximately 470~Mm$^2$, strictly avoiding network elements. For the $x$ direction we used cross correlation on the entire disk-part of the IRIS data excluding the slit in the slit-jaw images. To reduce projection effects we put the sub-FOV as far from the limb as the data allowed. Cross-correlation on two other, non-magnetic, sub-FOVs was done to estimate the uncertainty of the alignment. We performed the $x$ direction cross-correlation on the entire disk-part, including network, elements, as it gave overall better visual alignment compared to only using the sub-FOV. When using the entire disk-part for the $y$ direction, the computed shift in $y$ direction would be dominated by the magnetic network regions, for which TR line emission results in an ill-determined offset in the $y$ direction. This effect was not visible in the $x$ direction, likely because the network-based shifts canceled out because the network features are predominantly oriented along the $y$ axis, not the $x$ axis.

The \ca\ to 1600 alignment was performed manually. Shifts were applied every 285~s, and the shift of the intervening frames were computed from linear interpolation. We estimate the uncertainty to be within $0\farcs7$, approximately the same as the AIA resolution. We are not able to achieve sub-pixel accuracy because the images in the channels are not similar enough to compare large-scale patterns. We estimate that the uncertainty is the same for the \si\ to 1600 alignment. We remark that after numerous experiments we had to accept a relatively large uncertainty in the final alignment of the datasets. For on-disk multi telescope coordinated observations, we have been able to achieve much more accurate alignment to sub-pixel precision (e.g., \cite{pontieu_et_al_2014_science,martinez-sykora_et_al_2015,rouppe_et_al_2015}). For future coordinated limb observations we suggest the inclusion of a pure photospheric channel such as the IRIS 2832 SJI and the G-band filtergram on BFI/SOT.


We used a temporal grid with 19~s cadence and for each time selected the images closest in time in the different passbands. 19~s was chosen because that was the cadence of the \mg\ and \si\ data. For the time of each image we used the middle time of the exposure.

\subsection{Description of the FOV in the different filtergrams}
Figure~\ref{fig:fov} shows an overview of the limb part of the dataset with a dashed red line as a visual aid. For both datasets the red line is on top of a spicule. All other thin (less than 5~Mm wide) features seen sticking out are considered to be spicules. In the first dataset the red line is on top of a wide structure, wider than most spicules, and in the \ca\ and \si\ images we can see hints of substructure, multiple ``threads'', inside the large structure. All the ``threads'' move synchronously in time. \cite{skogsrud_et_al_2014} studied similar ``threads'' in higher resolution \ha\ data from the Swedish 1-m Solar Telescope and this may be the manifestation of those threads in the \ca\ and \si\ passbands.

Our general impression from working with these filtergrams is that the \mg\ images share characteristics with both the \si\ and \ca\ images, making them an excellent diagnostic for detecting spicules. The large scale structures are similar to the structures in the \si\ image, while many of the small scale features are similar to \ca\ structures, and the combination of \mg\ and \ca\ made it easier to find more spicules in the first dataset compared to the second dataset, which did not include \mg. Many spicules are barely observed in \si, for example in the region between $x=[20,40]$~Mm in the top example of Figure~1 where many clear spicules are visible in \ca\ and \mg.    

The \he\ image also appears to show many spicules, but the spatial resolution of AIA and potentially the optical depth are a clearly a limiting factor in resolving small scale structures. We also note that the average height at which the lowest point of spicules can be discerned is greatest in the optically thick lines of \he\ and \mg, while considerably lower in \ca\ and \si.

The spicule traced by the red dashed line in the first dataset is clearly visible in all images except \si, where it appears very faint. In contrast, the spicule sticking out to the right from the dashed line appears strongest in \si.

In the second dataset, which contains \carbon\ but not \mg, we observe that many spicules in \carbon, appear faint and close to the noise level in the data. There is little spicule substructure visible in the \carbon\ slit-jaw images.

\subsection{Radial filters}

\cah\ spicules, as observed by Hinode, are extremely faint and barely stand out in the original filtergrams. Radial density filters are then employed to enhance their visibility (e.g., \citealt{de_pontieu_et_al_2007b}; \citealt{zhang_et_al_2012}; \citealt{tiago_et_al_2012}; \citetalias{pereira_et_al_2014}). These filters are built by dividing each filtergram by $\left<I(r)\right>$, the normalized mean spatial and temporal intensity as a function of $r$, the distance from the limb. This greatly increases the contrast between the spicules and their background. However, $\left<I(r)\right>$ is mostly noise for large values of $r$, and dividing the filtergrams by such values becomes meaningless. Therefore, when building the radial filter $\left<I(r)\right>$ is scaled so that for large values of $r$ it tends to a constant value $\alpha$ that is higher than the background noise. The choice of $\alpha$ is largely subjective and determined by visual inspection, so that spicules are enhanced but artefacts and noise are minimized. In \citetalias{pereira_et_al_2014} and earlier work, $\alpha=0.135$ was used.

Spicules in the IRIS filtergrams are not as faint as in \cah, but radial filtering is still important to improve their visibility. In addition, as noted in \citetalias{pereira_et_al_2014}, spicule trajectories are easier to follow in IRIS than in \cah\ because they do not fade as much. The analysis of space-time diagrams in \citetalias{pereira_et_al_2014} showed that after fading, \ca\ spicules still showed an extremely faint outline that seemed to match the evolution of spicules in other filters. In this work we investigated this point further and found that this residual spicule signal in \ca\ filtergrams can be enhanced by using a more aggressive radial filter. Using $\alpha=0.018$ we get noisier images farther from the limb, but spicules are seen to extend farther out, to about the same heights as the \mg\ or \si\ spicules. The point of the radial filter is an important one because, as discussed in earlier literature, the height and lifetime measurements of spicules are critically dependent on their appearance as seen in filtergrams.

\begin{figure}
  \centering
  \includegraphics[width=0.99\columnwidth]{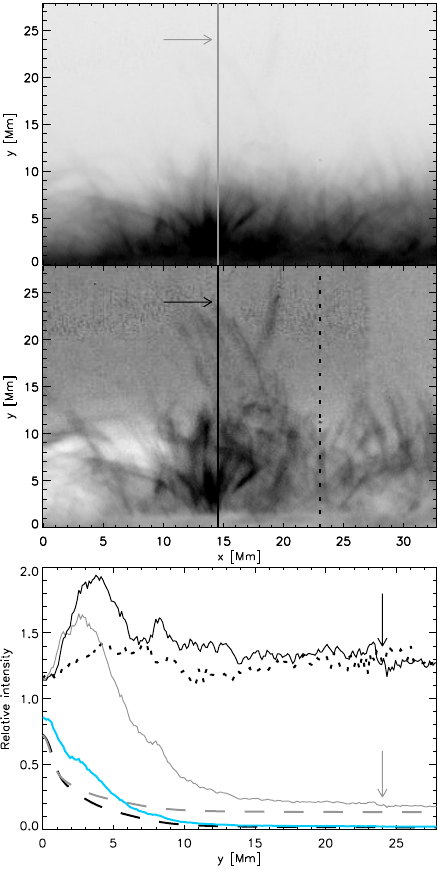}
  \caption{Comparison of radial filters. \emph{Top panel:} image with the radial filter used in \citetalias{pereira_et_al_2014} (inverted color table). \emph{Middle panel:} image with the radial filter used throughout this work. \emph{Bottom panel:} comparison of radial filters applied above. The dashed lines show the radial filter intensities $\left<I(r)\right>$ for the filters of \citetalias{pereira_et_al_2014} (gray) and this work (black). The solid gray/black lines show the intensities along the vertical lines of the same color in the panels above (gray: the radial filter of \citetalias{pereira_et_al_2014}, black: the radial filter of this work), and the solid blue line shows the original non-filtered intensities along the same vertical line. The dotted black line is the intensity along the dotted black line in the middle panel. The arrows indicate the top of a spicule.}
  \label{fig:radial}
\end{figure}

The two radial filters are compared in Figure~\ref{fig:radial}. The very faint spicules seen sticking out from about $(x,y)=(14,15)$~Mm toward the upper left of the image are not identifiable in the non-filtered intensity, and are barely discernible with the radial filter of \citetalias{pereira_et_al_2014} (see slight bump in intensity in the lower panel). However, they are seen more clearly with the new filter. By comparing with the intensity along the black dotted column (serving as an indicator of background intensity level, where no spicules exist at such heights), one can appreciate that the spicule signal is very low, comparable to the noise level (black dotted line). The temporal evolution of this spicule is shown in Figure~2 of \citetalias{pereira_et_al_2014}. Making a time series figure equivalent to their Figure 2 using our improved radial filter, we can trace the \ca\ spicule component for much longer: up to 456~s with the improved filter, compared to 248~s with the old filter.

We applied the improved radial filter to the \ca\ images. For the IRIS images there was no advantage in using this more aggressive filter, so we used the same radial filter as in \citetalias{pereira_et_al_2014}. For the AIA \he\ images no radial filter was applied. 


\subsection{Space-time plots}
The primary tool we use for analyzing the temporal evolution of spicules is space-time plots, \xt-plots, with $x$ being the distance along a slit and $t$ time. To obtain an \xt-plot of the spicules, a slit was manually drawn on the spicules for each time step for a 10~minute duration. This allows for the tracking of the transverse motion of spicules. The slits were drawn on the \ca\ and \mg\ images. Due to the alignment uncertainty, we need to take into account that the position of the features can be slightly shifted between the passbands. To accommodate these small shifts, a slit width of 1$\farcs$5 was used. For each height in the extended slit, the maximum intensity value was computed from the radial filtered images. We chose the maximum value instead of the average value to avoid the artifacts which then can occur. For example, if we move the artificial slit from one frame to the next to track a narrow spicule, and the spicule is perfectly centered in the first slit, but is slightly offset in the next, the average value will be shifted to a lower value because more of the dark background is included in the slit. It was verified that using a smaller or wider slit had little effect on the \xt-plots.  

The statistics for the spicules were obtained from the \xt-plots in all spectral passbands. 
The lifetime was defined to be the time difference between the onset and end of the visible spicule, i.e. the lifetime only measures the visible part of the spicule in the \xt-plots. The lifetime of a spicule differs in the different passbands and the lifetimes we measured are a lower estimate of the actual lifetime of the physical event: we can only trace isolated spicules above the ``spicule forest'' and most likely part of the lifetime of most spicules is spent hidden between other spicules. The minimum height at which the spicules are visible varies between the passbands, and between spicules due to interfering neighboring spicules. For example, in \ca\ the lowest positions of the spicules can often be seen around 4~Mm above the photospheric limb, while for \mg\ the lowest position is often seen at around 8~Mm above the limb. The way we detected spicules, by searching in all passbands, may give a bias toward the longer lived spicules. Our cadence of 19~s is also not sufficient to detect spicules with lifetimes of less than about a minute. In \cite{tiago_et_al_2012}, the lifetimes were determined based on filtergram inspection. It is possible that by using \xt-plots instead, with the knowledge of the spicule position after it fades in \ca, that a bias toward longer lifetimes is introduced because the faint trails left after the spicule fades might not be associated with the spicule in $xy$-snapshots at the time.

Our length measurement follows the spicule from the top, at time of maximum extent, down toward the limb to the point where the line-of-sight superposition, ``spicule forest'', obscures the spicule. From that point the shortest line to the photospheric limb is added. If the spicule has an angle to the limb normal, or the actual footpoint is below the limb due to the projection in the observations, we are underestimating its length.

\begin{figure}
  \centering
  \includegraphics[width=0.99\columnwidth]{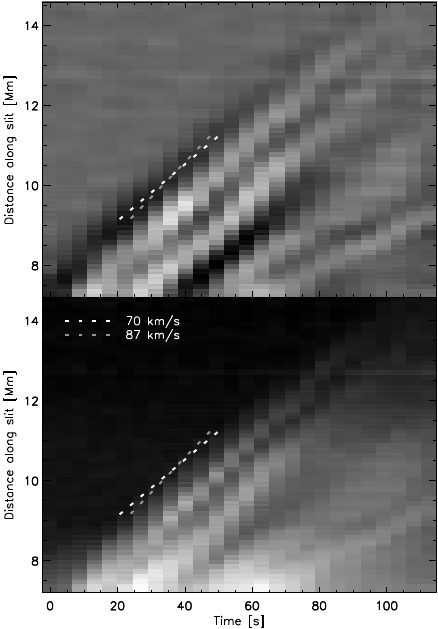}
  \caption{Illustration of how the peak ascent velocity was extracted from the \ca\ \xt-plots. The upper panel shows an edge-enhanced version of the \ca\ image in the bottom panel. The two dashed lines in each panel both represent an equally good estimate of the velocity (values printed in the legend).}
  \label{fig:peakvel}
\end{figure}

We measured the peak ascent velocity by fitting a straight line to the steepest part of the \xt-plot. We applied an edge enhancement algorithm beforehand to the \ca\ \xt-plot to make the manual extraction easier for the spicules that did not have a clearly defined upper point. Because there is a large uncertainty in the manual approach, we did the measurements twice to estimate the consistency of our results. The average difference between the two measurements was -2~\kms\ with a standard deviation of 21~\kms. Figure~\ref{fig:peakvel} illustrates how the velocities were extracted with two different good fits with about 17~\kms difference between them.

\section{Results}
We measured a total of 54 spicules in the two datasets, 35 in the first and 19 in the second. The large difference in number of spicules detected is because the second dataset was of shorter duration, but also because \mg\ was not observed, which made cross-channel identification harder (see Section~\ref{sec:observations}).

\begin{figure*}
  \centering
  \includegraphics[width=0.99\textwidth]{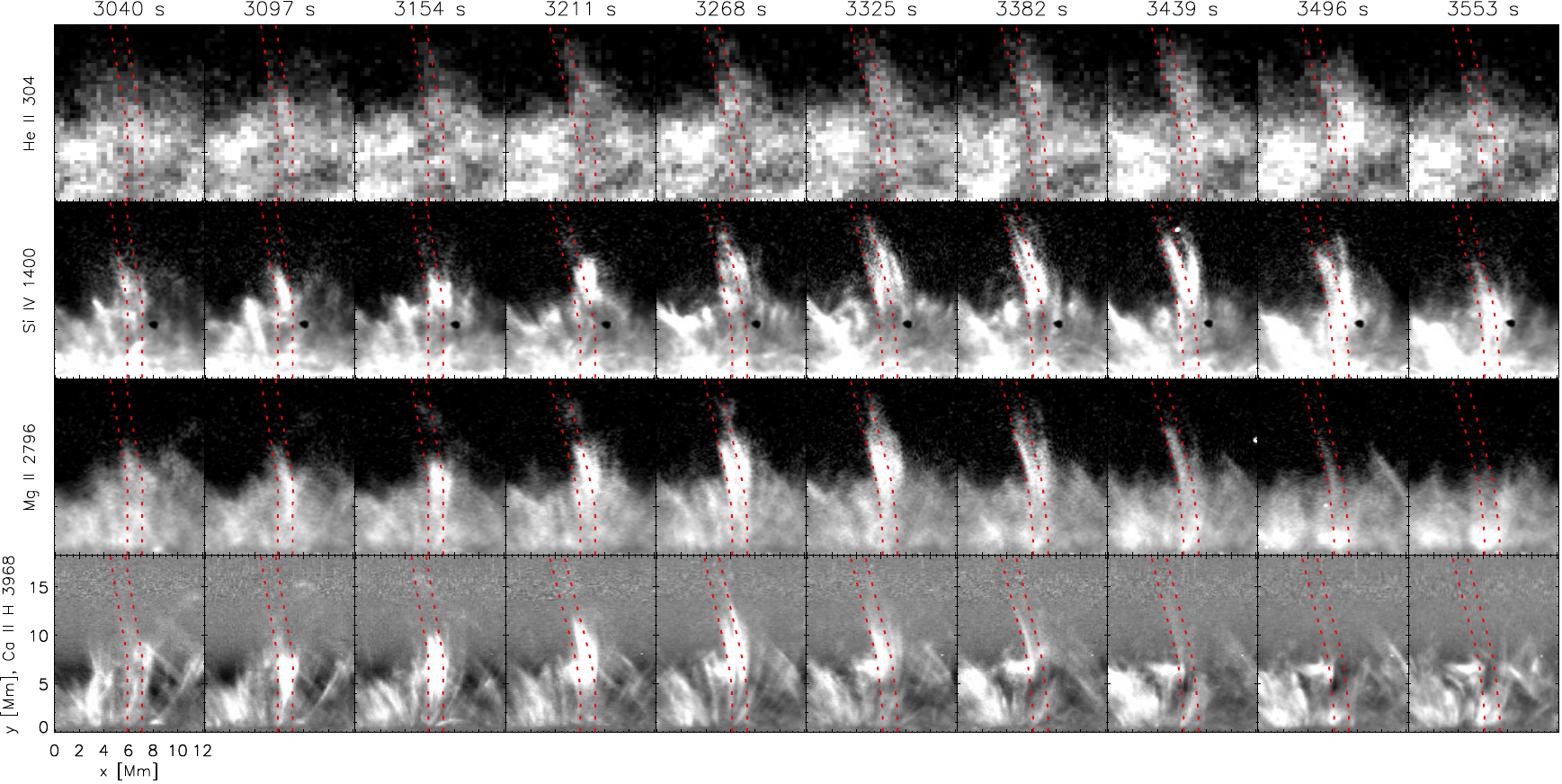}
  \includegraphics[width=0.99\textwidth]{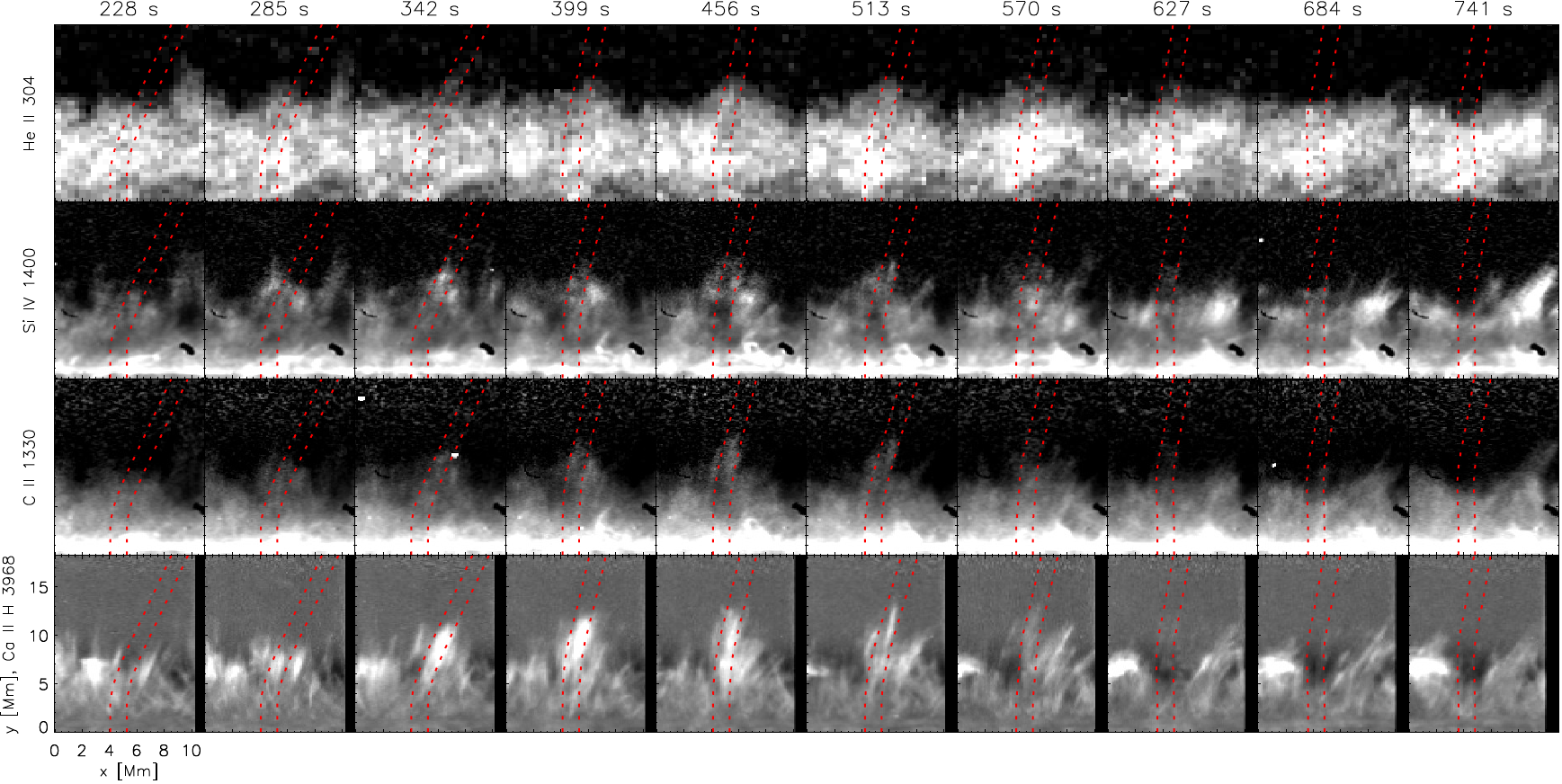}
  \caption{Time series of two spicules, one from each dataset. The spatial cutouts are fixed in space and the artificial slit used for the construction of the \xt-diagrams is between the two red dashed lines. The corresponding \xt-plot to the top spicule is \#1 in Figure~\ref{fig:xt} and \#4 for the bottom spicule. Note that all images except \he\ are radially filtered.}
  \label{fig:tseries}
\end{figure*}

In Figure~\ref{fig:tseries} we show the time evolution of two spicules, one from each dataset. The first spicule appears to have significant substructure and is seen rising and falling in all passbands except \ca, where it fades significantly. The slit is centered on the left-most of the two substructure ``threads''. The length of the spicule is comparable in all passbands. For the second spicule the length is also comparable in the passbands. The spicules fade in the downfall phase in the \ca, \carbon\ and \si\ passbands. The \carbon\ component appears very faint.

In \citetalias{pereira_et_al_2014}, three spicules are shown in detail and all are clearly detected in \ca, \mg, \si\ and \he. From the spicules we measured in the same dataset, we find that 23 out of 35 spicules (66\%) have a clear component in all four passbands. Of the 12 that do not have a clear component in all passbands, the \si\ component is unclear in 10 and the \he\ component in 2. By unclear we mean that the relatively thin structure of the spicule seen in the other passbands is not identifiable, instead we observe emission from a much larger structure, possibly obscuring the spicule. Of the 54 \ca\ spicules we analyze we find that 41 (76\%) fade rapidly, usually in less then 38~s (i.e., in two timesteps). From those that fade, 22 (54\%) show a faint trail in the \xt-plots at the location of the spicule in passbands other than \ca. These faint leftover traces are not easily connected to the earlier spicule phase if we were not aided by knowledge of the position of the spicule in the other passbands, see also \citetalias{pereira_et_al_2014}. Of the \mg\ spicules, 11 out of 35 (31\%) are seen to fade, 3 out of 36 (8\%) \si\ spicules fade, 1 out of 19 (5\%) \carbon\ spicules fade and only 1 out of 54 (2\%) \he\ spicules are seen to fade. \mg\ spicules typically fade over a height range between 3-10~Mm in less than 38~s. We observe that if a spicule fades from a passband, it generally also fades from the passbands with a lower formation temperature. For example, a spicule can fade from the \mg\ passband and still be visible in the \si\ passband, but not the other way around.

\begin{figure*}
  \centering
  \includegraphics[width=0.99\textwidth]{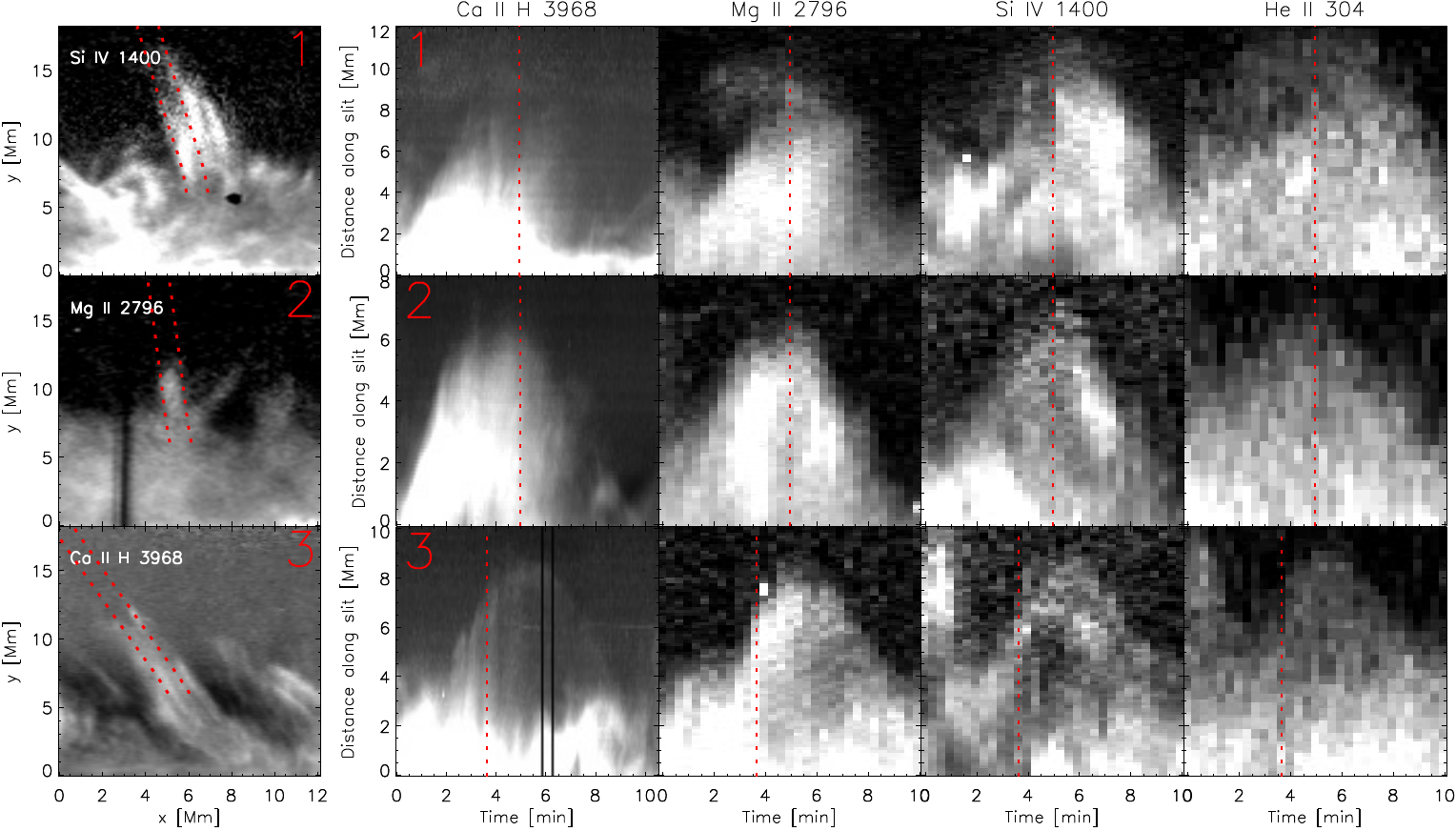}
  \includegraphics[width=0.99\textwidth]{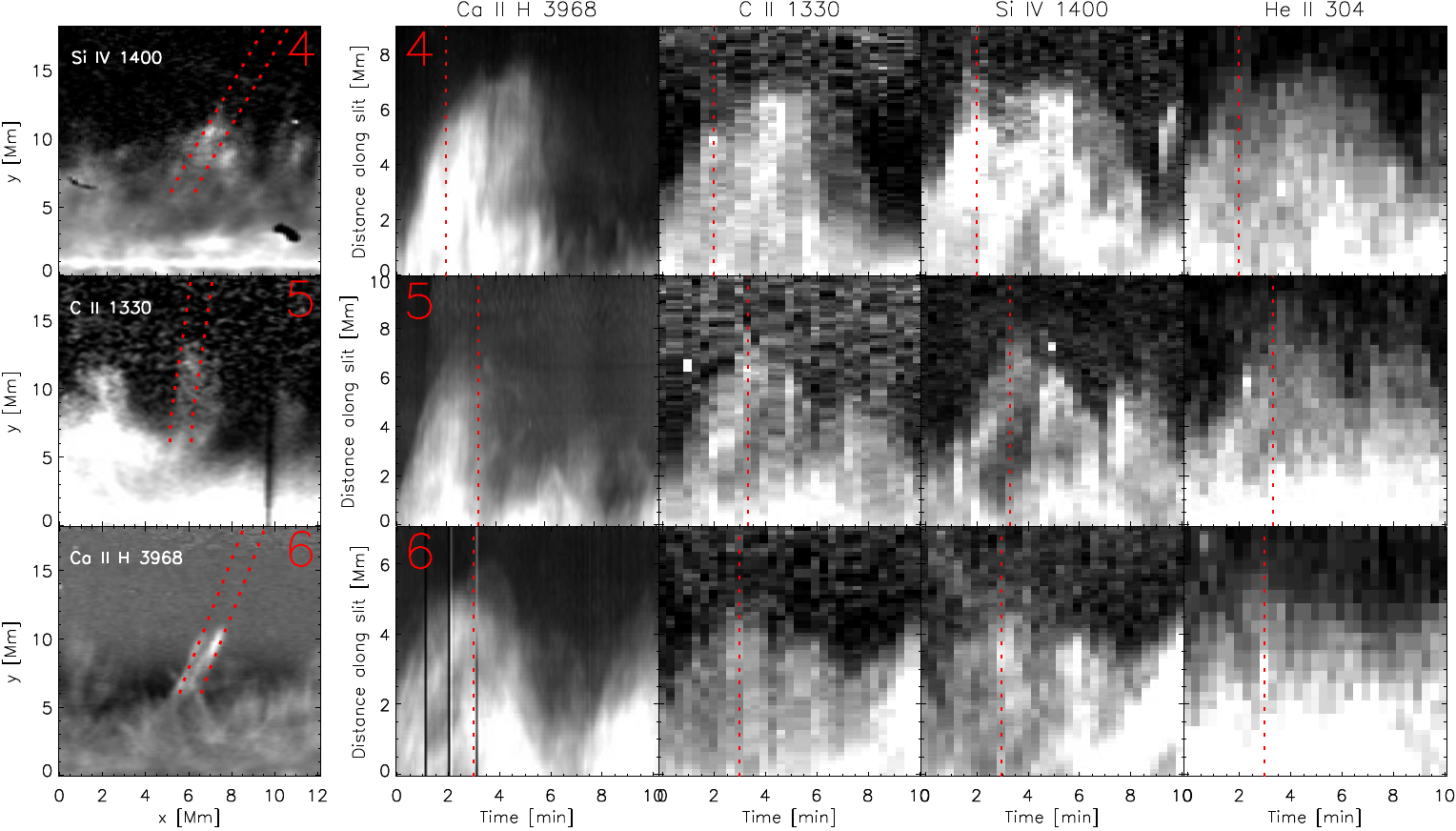}
  \caption{$xt$--diagrams of six spicules from the two datasets. The first column is a spatial cutout of the FOV at the time marked with a red dashed line in the corresponding $xt$--plots on the same row. The slit regionis between the two red dashed lines. The black and white stripes visible in \ca\ for spicules \#3 and \#6 are due to missing data. Note that all images except \he\ are radially filtered. Time series of spicule \#1 and \#4 are shown in Figure~\ref{fig:tseries}. Animations of this figure is available in the online material. \textbf{(Movies here: \href{http://folk.uio.no/haaksk/movies/xt-1.mp4}{movie-1}, \href{http://folk.uio.no/haaksk/movies/xt-2.mp4}{movie-2}, \href{http://folk.uio.no/haaksk/movies/xt-3.mp4}{movie-3}, \href{http://folk.uio.no/haaksk/movies/xt-4.mp4}{movie-4}, \href{http://folk.uio.no/haaksk/movies/xt-5.mp4}{movie-5}, \href{http://folk.uio.no/haaksk/movies/xt-6.mp4}{movie-6}}}
  \label{fig:xt}
\end{figure*}

In Figure~\ref{fig:xt} we show \xt-plots for six different spicules selected to illustrate the typical behavior seen in our data. For each spicule we show a snapshot, in different passbands, from the time given by the vertical dotted red line in the \xt-plots on the same row.  The bottom three spicules are from the second dataset, containing \carbon. For spicules \#1-\#5 there is a visible component in all passbands, and for \#6 only the \ca, \carbon\ and \si\ components are clearly visible. For all the \ca\ spicules we observe that the final phase is significantly weaker compared to the onset phase, \#3 and \#4 are very clear examples. In spicule \#3 we observe emission from only the top of the spicule in \ca\ during downward phase after it has faded. The path traced by the top matches the parabola seen in the other passbands. For the \si\ components we see that \#1-\#3 and \#5 show an increase in emission in the final phase and in \#2 and \#5 the \si\ component is brighter toward the top. We find that 24 of the 54 (44\%) \si\ spicules show increased emission in the top part of the spicule. 

In \citetalias{pereira_et_al_2014} it is reported that, for some examples, the \si\ emission is increased when the \ca\ spicule component fades. We find that in 30 out of the 54 spicules (56\%) the \si\ channel displays increased emission when the \ca\ components fade. Spicules \#1-\#3 and \#5 in Figure~\ref{fig:xt} are clear examples of this behavior. In \#1 and \#2 we see that the \si\ component brightens up when the \ca\ components fade, and that the \si\ \xt-plot is a natural (parabolic) continuation of the \ca\ \xt-diagram. By looking at the evolution of intensities we estimate the delay between the channels to be about 3~minutes. In \#3 and \#5 we observe that more of the middle part of the \si\ spicule brightens up below the top point when the \ca\ component disappears. Table~\ref{tab:dat} summarizes selected results. 

\begin{figure}
  \centering
  \includegraphics[width=\columnwidth]{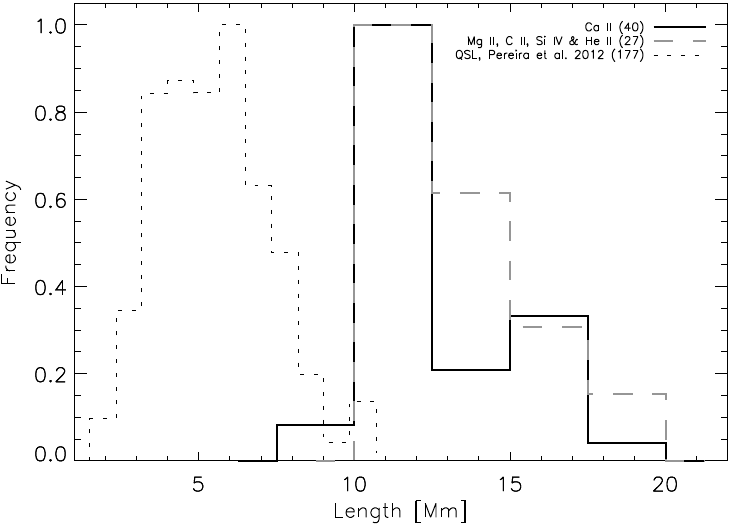}
  \includegraphics[width=\columnwidth]{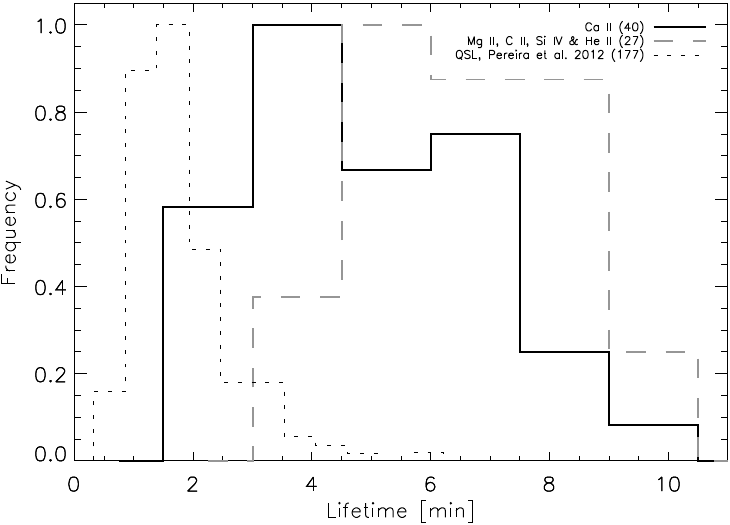}
  \includegraphics[width=\columnwidth]{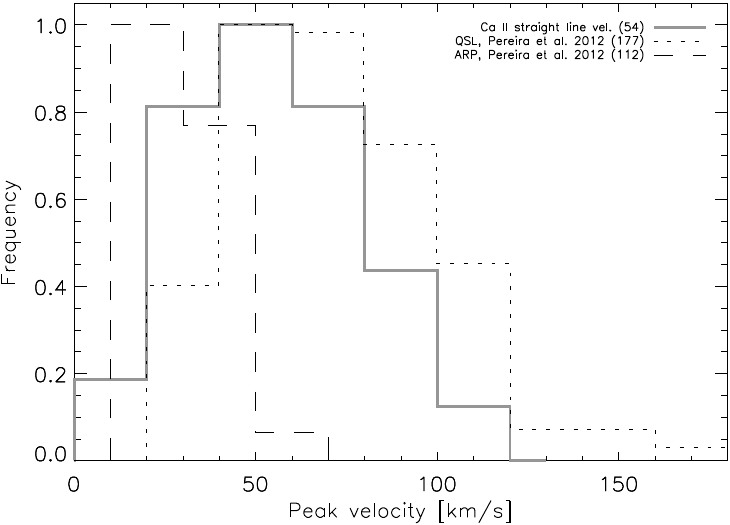}
  \caption{Comparison of our results with previously reported statistics for quiet Sun linear spicules (QSL), type II spicules, in \ca. In the bottom panel we included reported velocities from active region parabola spicules (ARP), type I spicules. Note that the ARP velocities were computed from the first derivative of the parabolic fit and the other velocities are from fitting a straight line to the \xt-plots.}
  \label{fig:hist}
\end{figure}

\begin{figure}[!htb]
  \centering
  \includegraphics[width=0.99\columnwidth]{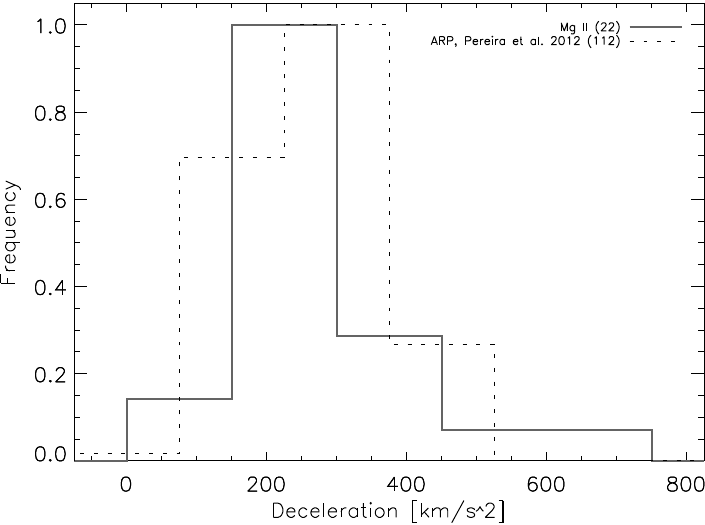}
  \caption{Comparison of our measured \mg\ decelerations with previously reported decelerations for active region parabola spicules (ARP), type I spicules, in \ca.}
  \label{fig:hist2}
\end{figure}

It is of interest to compare the statistics of our \ca\ data with the existing numbers reported for type II spicules in \ca. In Figure~\ref{fig:hist} we compare our measured spicule properties with the results of \cite{tiago_et_al_2012} for quiet Sun spicules in \ca\ data. The figure shows histograms of the maximum height, peak ascent velocity and lifetime. The lengths we measure are between 8-20~Mm with 12.5~Mm average for \ca\ and 13~Mm average for \mg. The difference in results between the \hinode, IRIS and AIA passbands are small. In the lifetime plot we can see that the \ca\ lifetimes are slightly shifted toward shorter duration. This is what we expect based on the observation that more spicules fade from \ca\ compared to the other passbands. The \ca\ average lifetime is 5~min, while for the remaining passbands it is about 6.5~min. The longest lifetime found is 10.5~min. The average peak velocity we measure is 53~\kms\ and the maximum velocity we find is 108~\kms. We find that the lengths and lifetimes we measure are significantly longer than previously reported. The velocities match well for quiet Sun spicules, and are significantly higher than reported type I spicule velocities.

In Figure~\ref{fig:hist2} we compare our measured \mg\ decelerations with previously reported numbers for active region parabola spicules, identified as type I spicules. The decelerations match well with reported type I decelerations. We determined the deceleration from fitting a parabola to the \mg\ \xt-plot. The reason for choosing \mg\ is that the \ca\ and \si\ \xt-plots often show an incomplete parabola, while the \mg\ shows a complete parabola. We performed the measurements on a subset of 22 \mg\ spicules that showed clear and complete parabolas in the \xt-plot for its entire duration. 

\begin{deluxetable}{lccc}
\centering
\tablecaption{spicule evolution statistics \label{tab:dat}}
\tablehead{\colhead{Element} & \colhead{\# spics} & \colhead{\% that fade}  & \colhead{\% show parabolic \xt-path} } 
\startdata
\ca     &  54 & 76 &  33 \\
\mg     &  35 & 31 &  82 \\
\carbon &  19 & 5  &  64 \\
\si     &  54 & 8  &  68  \\
\he     &  54 & 2  &  62 
\enddata
\end{deluxetable}
\vspace{0.1cm}

\section{Discussion}
We studied the detailed evolution of 54 spicules in the \cah, \mglong, \silong, \carbonlong\ and \helong\ filtergrams. We find that a typical spicule, in quiet Sun regions, has a visible component in all filtergrams. The \ca\ component typically fades away and the evolution continues in the other passbands. Most \ca\ spicules that fade leave a faint ``trace'' in space-time plots. We find that 44\% of the \si\ spicules are brighter toward the top, while 56~\% of the spicules show an increase in \si\ emission when the \ca\ component fades. We interpret this behavior as the effect of heating in the spicule, which is consistent with earlier work (\citealt{de_pontieu_et_al_2007b}; \citealt{de_pontieu_et_al_2011}; \citealt{tiago_et_al_2012}; \citetalias{pereira_et_al_2014}; \citealt{rouppe_et_al_2015}).

In \citetalias{pereira_et_al_2014} it was found that \mg\ and \si\ spicules continue to rise for 4-8~Mm after the \ca\ spicule component fades and that the \mg\ and \si\ spicule lifetimes were longer by several minutes. In this study, using a more aggressive radial filter on the \ca\ data, we find that \ca\ spicules are more similar to \mg\ spicules than it appeared in \citetalias{pereira_et_al_2014}. We are pushing the \ca\ data to the noise limit and exploiting the access to \mg\ data to detect very faint spicule signal that was otherwise indiscernible. We observe much more similar heights and lifetimes for \ca\ and \mg\ spicules. This is what we expect because the elements have roughly the same formation temperature. The small difference in lifetimes we find between \ca\ and \mg\ are perhaps due to the different opacities of the spectral lines, Mg is about 18 times more abundant than Ca \citep{asplund_et_al_2009}. However, because the intensity ratio of \ca\ and \mg\ is not constant as function of time there must be other effects that also play a role.

We find several examples of spicules that fade from passbands other than \ca, and we note that if a spicule fades from a passband, it also generally fades from the passbands with lower average formation temperature. We find more spicules that fade from the cooler passbands than from the hotter passbands. 

When comparing our results to \ca\ quiet Sun spicules of \cite{tiago_et_al_2012} we find good agreement in peak ascent velocity. When comparing our results to type I spicules, the active region parabola spicules of \cite{tiago_et_al_2012}, we find similar decelerations for our \mg\ spicules as for \ca\ type I spicules, but we find much higher peak velocities in the ascent phase than in \ca\ type I spicules. 

It is important to discuss what these new multi wavelength results on spicules mean for our understanding of the difference between type I and II spicules. Now that we find clear parabolic paths in the \mg\ and higher temperature passbands and at least traces of parabolic paths in part of the \ca\ spicules, the question arises whether distinction between types of spicules is still justified. The spicules we study have higher peak velocity compared to type I spicule, and they clearly undergo heating, which is not compatible with type I spicules. The longer lifetimes and heights we can extract from the data do not alter the fundamental aspect of type II spicules, that they fade in the different passbands and particularly in the cooler (or lower opacity) passbands. We observe that the \ca\ spicules fade significantly and that the downward phase is primarily only visible in the IRIS and AIA data. The parabolic paths in the \ca\ data only become apparent after aggressive filtering and it pushes the data to the limits fundamentally set by noise.

It is now clear that the spicules we study are subject of heating. If we assume that ionization equilibrium is valid in these dynamic events, then the standard estimated temperature ranges for the different passbands would be correct, and we would conclude that we always observe spicules to consist of a mix of chromospheric and transition region temperatures. However, the dynamic nature of these jets indicates that non-equilibrium ionization may well play an important role, which would change our interpretation of what occurs in spicules. Regardless of the ionization state, our observations support a scenario in which heating is occurring, often toward the top of the spicule. We measure typical timescales between peak intensity in \ca\ and \si\ to be about 3~minutes. This can be interpreted as a typical time scale for the heating process, although it is possible that non-equilibrium ionization effects could change this value.

In our space-time plots we generally observe parabolic motions. The \ca\ \xt-diagrams often show only the early, more linear, phase of the parabola. Parabolic paths can be the result of driving by shock waves \citep{hansteen_et_al_2006, heggland_et_al_2007, heggland_et_al_2011}. However, it is difficult to envision a scenario in which magneto-acoustic shocks (of the type seen in \cite{hansteen_et_al_2006}) could lead to observations of heating like in our observations. In shocks the heating is occurring at the shock front as it passes through the material, and the aftermath of the shock is responsible for the parabolic motion. This does not seem to be compatible with our observations. The \mg\ parabolas are generally the most complete, unlike in \ca\, which often only shows the beginning, and in \si\, which is sometimes weak in the beginning and shows increased emission in the later stages. One can speculate whether this is an optical depth effect: possibly the spicule is optically thick in \mg, but optically thin in \ca\ and \si. Heating of the spicule, not necessarily uniformly across and along the whole spicule structure (e.g., in a subset of threads), might result in fading from the optically thinner \ca\ while there is still sufficient opacity in \mg\ in cooler parts of the spicule. Only the hotter part of the spicular plasma can then be seen in \si. This observational puzzle puts strong constraints on numerical models.

The increased emission we see in the \si\ \xt-plots, often occurring in the downward phase of the spicules, is compatible with previous on-disk observations of an apparent correlation between red-shifted H-alpha spicular plasma and C IV emission \citep{de_wijn_et_al_2006}. It also makes one wonder what role spicules play in the pervasive redshifts observed in transition region lines. This warrants further investigation.

\acknowledgements

IRIS is a NASA Small Explorer mission developed and operated by LMSAL with mission operations executed at NASA ARC and major contributions to downlink communications funded by the NSC (Norway). \hinode\ is a Japanese mission developed by ISAS/JAXA, with the NAOJ as domestic partner and NASA and STFC (UK) as international partners. It is operated in cooperation with ESA and NSC (Norway). The research leading to these results has received funding from both the Research Council of Norway and the European Research Council under the European Union's Seventh Framework Programme (FP7/2007-2013) / ERC grant agreement nr. 291058. B.D.P. is supported by NASA contract NNG09FA40C (IRIS), and NASA grants NNX11AN98G and NNM12AB40P. 

\bibliographystyle{apj}

\end{document}